\documentclass[a4paper,twocolumn,11pt]{quantumarticle}
\pdfoutput=1
\usepackage[utf8]{inputenc}
\usepackage[english]{babel}
\usepackage[T1]{fontenc}
\usepackage{amsmath}
\usepackage{hyperref}

\usepackage{tikz}
\usepackage{lipsum}
\newcommand{\black}{\color{black}}

\newcommand{\be}{\begin{equation}}
\newcommand{\ee}{\end{equation}}

\newcommand{\bea}{\begin{eqnarray}}
\newcommand{\eea}{\end{eqnarray}}
\newcommand{\gsim}{\lower.7ex\hbox{$\;\stackrel{\textstyle>}{\sim}\;$}}
\newcommand{\lsim}{\lower.7ex\hbox{$\;\stackrel{\textstyle<}{\sim}\;$}}

\usepackage[utf8]{inputenc} % Acentos
\usepackage{multirow} % para las tablas
\usepackage{float} % para comando \begin{table}[H]
\usepackage{ragged2e} %\justifying

\newcommand{\beq}{\begin{equation}}
\newcommand{\eeq}{\end{equation}}

\begin{document}

\title{Interaction-free measurements and counterfactual computation in IBM quantum computers
}

\vspace{.5cm}

\author{J. Alberto Casas} \thanks{j.alberto.casas@gmail.com}\orcid{0000-0001-5538-1398}
\affiliation{Instituto de F\'{\i}sica Te\'orica, IFT-CSIC/UAM, Universidad Autónoma de Madrid, Cantoblanco, Madrid, Spain}

\author{Bryan Zaldivar}\orcid{0000-0002-6313-6525}
\thanks{bryan.zaldivarm@uam.es}
\affiliation{Instituto de F\'{\i}sica Te\'orica, IFT-CSIC/UAM, Universidad Autónoma de Madrid, Cantoblanco, Madrid, Spain}
\affiliation{Departamento de  F\'{\i}sica Te\'orica, Universidad Autónoma de Madrid, Cantoblanco, Madrid, Spain}

\bigskip\bigskip\bigskip\bigskip

%\twocolumn[
%\begin{@twocolumnfalse}
	%\maketitle
	\begin{abstract} \normalsize
	
		The  possibility  of  interaction-free  measurements  and  counterfactual  computations  is  a  striking feature of quantum mechanics pointed out around 20 years ago. We implement such phenomena in actual 5-qubit, 15-qubit and 20-qubit IBM quantum computers by means of simple quantum circuits. The results are in general close to the theoretical expectations. For the larger circuits (with numerous gates and consequently larger errors) we implement a simple error mitigation procedure which improve appreciably the performance.

	\end{abstract}
	%\begin{center} \vspace*{0.2cm} \end{center}
%\end{@twocolumnfalse}
%]

\maketitle

\vskip 0.2cm

\setlength\parskip{0.1cm}

\section{Introduction}

\begin{figure}[htb]
\begin{center}
\includegraphics[height=5   cm,clip=]{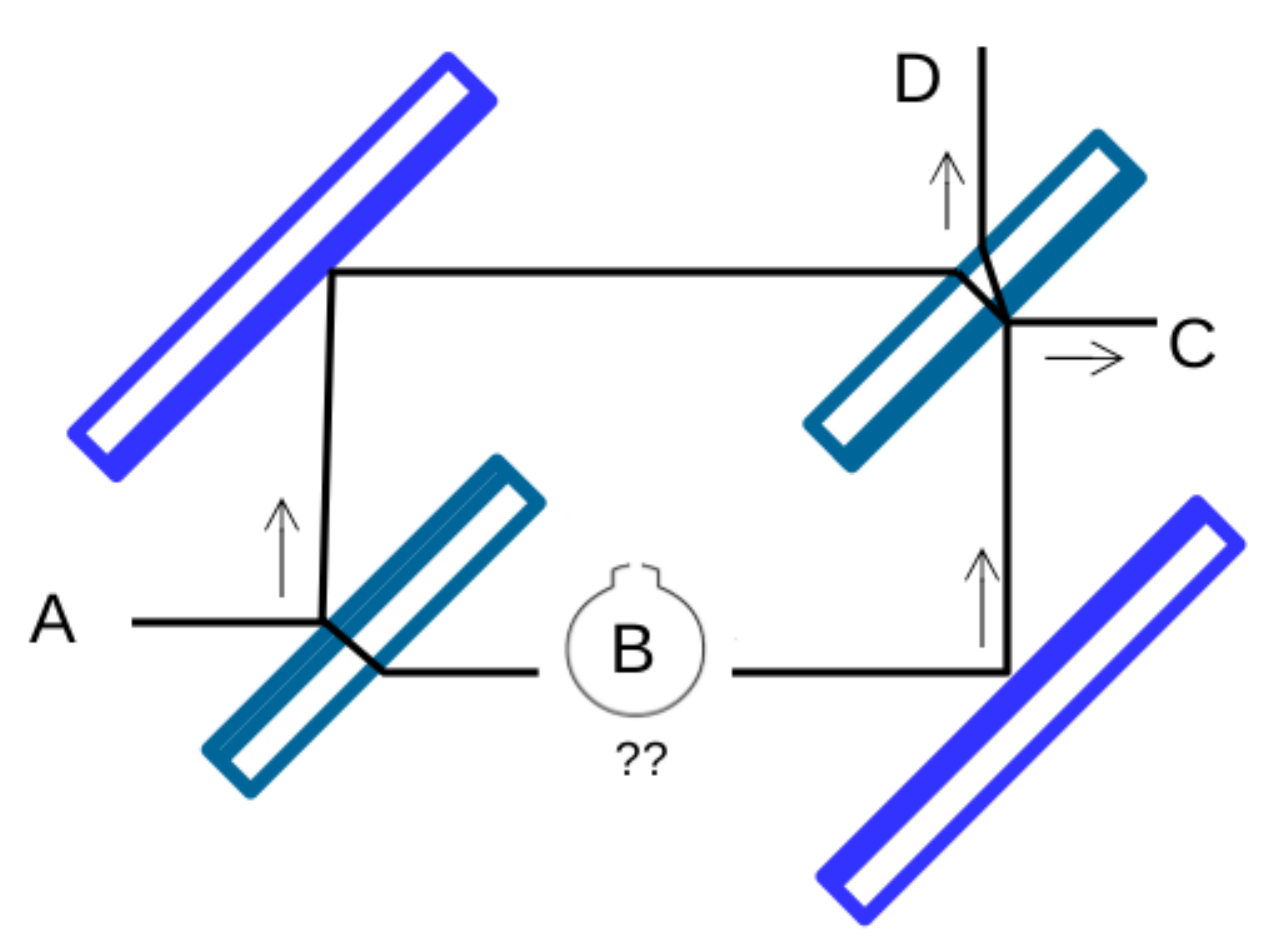} 
\caption{Elitzur and Vaidman bomb tester. 
}
\label{fig:bomb}
\end{center}
\end{figure}

A fascinating feature of quantum mechanics is the possibility of realizing interaction-free measurements, in which non-trivial information about a system is obtained without disturbing it. They are also called 
{\em counterfactual}, to highlight the fact that one is exploring ``what would have happened if...", without actually happening.
This concept was first introduced by Elitzur and Vaidam \cite{Elitzur:1993xh} and experimentally demonstrated by Kwiat et al. \cite{Kwiat:1995}. 
Specifically, the original idea of the gedanken experiment was to select a bomb (without destroying it) from a supply of bombs (some of which are duds) that would explode when detonated by a photon impacting its sensor (the duds have no sensor); an impossible task on classical grounds. To that end, Elitzur and Vaidam conceived a devise consisting of a Mach-Zehnder interferometer, placing the bomb in one of the arms, Fig. \ref{fig:bomb}. Then a single photon is %launched
emitted (from point $A$ in Fig. \ref{fig:bomb}), entering a superposition after passing through the first beam splitter. In the absence of bomb, or if the bomb is a dud, the two paths of the photon interfere at the last semi-transparent mirror in a constructive (destructive) way along the direction towards detector C (D). Thus the photon ends up at C. However, if the bomb sensor works, it acts as a measuring device. Half of the times the photon will collapse at the bomb, which would explode. The other half the photon collapses at the upper arm. Since the superposition is destroyed, the surviving photon will end up at detectors C and D with equal probability. In other words, if the bomb works (does not work), there is a 25\% (0\%) probability that the photon arrives at detector D. In that case the bomb is selected without any damage.

Later, Jozsa, and Mitchison and Jozsa \cite{Jozsa:1998bq, Mitchison:1999uf} applied this idea to show the theoretical possibility of counterfactual computations, i.e. instances in which a (simple) computation is realized with the computer switched off. They offer a particularly simple example of this. Suppose that the computer (more realistically, a logic gate) implements a 1-bit to 1-bit function, $f_r$ (unknown to us), which acts on the bit 0 as $f_r(0)=r$ (with $r = 0$ or 1), and we wish to determine the value of $r$ without actually switching on the gate.
The relevant part of the system is described by two qubits, $|ab\rangle$. The first one acts as the switch that controls the computer: $a=0\ (1)$ for the computer switched-off (on). The second one is the register qubit for the input/output. Before the calculation the system is at the initial state $|10\rangle$ ($|00\rangle$) if the computer is turned on (off). Then, after the time needed for the calculation the state becomes $|1r\rangle$ if the computer was switched-on, or it remains unchanged, $|00\rangle$, if it was switched-off. 

The protocol for a counterfactual computation devised by Jozsa \cite{Jozsa:1998bq} gives the possibility to obtain the result of the calculation when this is $r=1$ without ever switching on the computer:
%
%{\blu 
%The computer (more realistically, a logic gate) implements a 1 bit to 1 bit function $U_r (|{\rm input}\rangle)$, and we wish to know the value of the parameter "r", which for simplicity here is binary (i.e. $r = 0$ or 1), without actually performing the computation. 
%In order to do that, they design a protocol containing two qubits, the first one acting as a switch to turn on/off the computation, and the second one being the actual input of the computation. The switch on (off) controlled by the first qubit is analogous to the photon passing (not passing) through the lower arm of the bomb tester, where the bomb is located. On the other hand, the value of $r=1 (0)$ is analogous to the bomb being live (dud). Such protocol, described below in more detail, will give as an output the knowledge that $r=1$ with the switch turned off; in other words, it will allows us to learn such a result in a counterfactual way, in the analogous way to the bomb tester thought experiment, by which we can know whether the bomb is live without actually activating it.}

%
\begin{enumerate}

\item
Start with initial state 
\be
|\psi_{\rm in}\rangle=|00\rangle\ ,
\ee
i.e. with the `computer' switched off and the input at 0.

\item Perform a unitary transformation in the switch qubit, rotating it an angle $\theta=\frac{\pi}{2N}$. The new state becomes
\be
|\psi_1\rangle=\cos \theta|00\rangle +\sin\theta|10\rangle\ ,
\label{theta_rot}
\ee
i.e. the switch is in an off-on superposition.

\item Let the system to evolve a time long enough for the calculation to be performed in the computer. The state becomes
\be
|\psi_1'\rangle=\cos \theta|00\rangle +\sin\theta|1r\rangle
\ee
Note that for $r=0$, $|\psi_1'\rangle=|\psi_1\rangle$.
\item 
Measure the second qubit in the computational basis. If $r=0$ the result of the measurement is '0' with probability 1, and the state remains unchanged, i.e. $|\psi_1\rangle$. If $r=1$, there is a  $\cos^2\theta$ ($\sin^2\theta$) chance that the result is 0 (1); then the state of the system collapses into $|00\rangle$ ($|11\rangle$). 

Note that, for $r=1$, if the result of the previous measurement were
1
the computer has been switched on, and the method has failed (though we have learnt that $r=1$). If it was 0, the computer remains switched-off. 

\item 

Repeat steps 2-4 $N$ times in total. If $r=1$, there is a global probability $(\cos^2\theta)^{N}$ that the final state is $|00\rangle$; if $r=0$ at each iteration the state rotates an angle $\theta$, so at the end of the process it becomes $\cos (N\theta)|00\rangle +\sin(N\theta)|10\rangle = |10\rangle$. 

{\black
\item {\black Measure the first qubit. If $r=0$, the measurement will yield 1 with probability 1. If $r=1$ it will yield 0.}}

%Repeat steps 2-4 $N$ times in total. If $r=1$, there is a global probability $(\cos^2\theta)^{N}$ that the final state is $|00\rangle$. If $r=0$ at each iteration the state rotates an angle $\theta$, so at the end of the process it becomes $\cos (N\theta)|00\rangle +\sin(N\theta)|10\rangle = |10\rangle$.

\end{enumerate}
Therefore, if $r=1$ there is a global probability $(\cos^2\theta)^{N}$, which tends to 1 for large $N$, to determine the result of the computation with the computer switched off, i.e. in a counterfactual way. In contrast, if $r=0$, the computer has been switched on. Mitchison and Jozsa have argued that in any quantum protocol the sum of the probabilities to get both $r=0$ and $r=1$ in a counterfactual way cannot be larger than 1; so this example, in the large $N$ limit, saturates the theoretical bound\footnote{The Mitchison and Jozsa bound has been discussed in refs.\cite{Hosten:2006, Vaidman:2006, MJ:2006}}.

The aim of this work is to implement interaction-free measurements and counterfactual computations, as the one described above, 
in the quantum computers of IBM Quantum Experience,
by using simple quantum circuits.

\section{Interaction-free measurements in a quantum computer}

    In ref.\cite{Das} Das et al. have designed a quantum circuit to somehow mimic the architecture of the Elitzur and Vaidman bomb tester idea ((for earlier work in the subject see Refs.\cite{Paraoanu, Zhou})), representing the photon direction by a pair of qubits and using combinations of gates to represent the beam splitters and mirrors involved in the Mach-Zehnder interferometer; in addition to those to mimic the bomb. However, although the circuit may be a fair representation of the quantum bomb tester, the proliferation of gates is expected to induce large deviations from the theoretical result when the circuit is ran in a real quantum computer\footnote{We have verified this in the ibmqx2 5-bit quantum computer \cite{IBMQ}. Namely, about 14\% of the outputs (instead of the theoretical 0\%) correspond to a state that has no interpretation in that context.}. 

\begin{figure}[htb]
\begin{center}
\includegraphics[height=3cm,clip=]{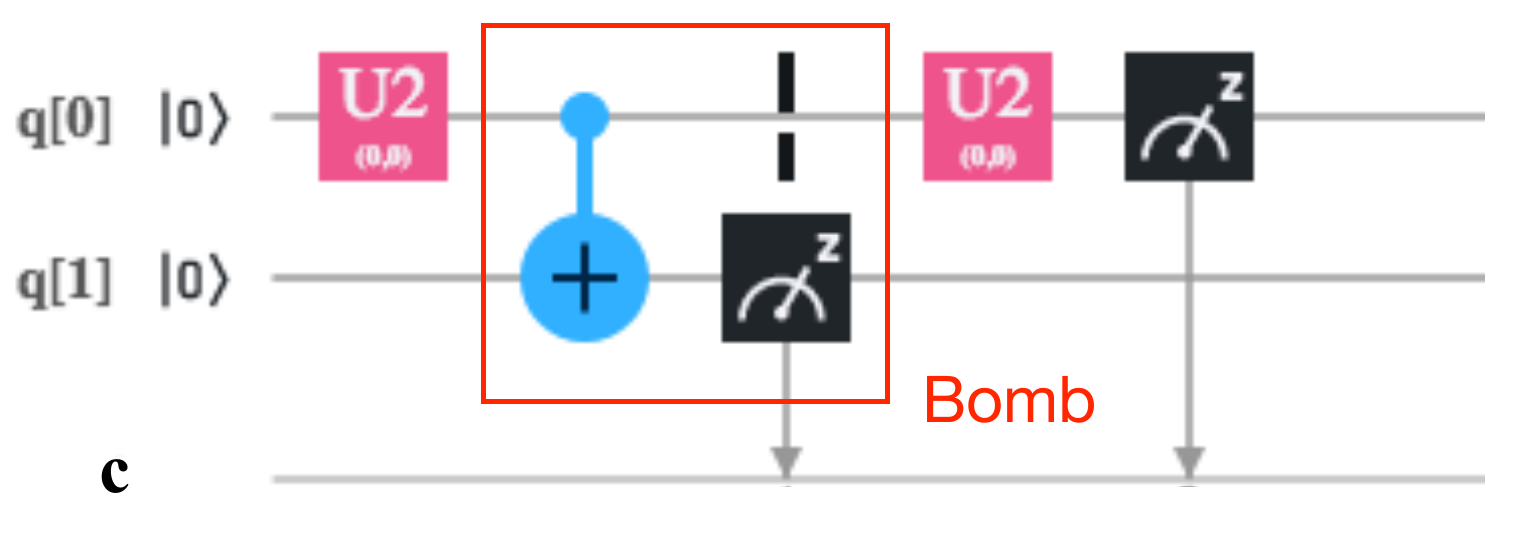} 
\caption{Bomb tester circuit}
\label{fig:bombtester}
\end{center}
\end{figure}

On the other hand, it is in fact quite easy to implement the Elitzur and Vaidman bomb tester idea by means of the simple quantum circuit of Fig. \ref{fig:bombtester}. The bomb is represented by a CNOT gate (which plays the role of the sensor) followed by a measurement unit, which represents the bomb explosion when the result is 1. The control qubit, $q0$, corresponds to the switch of (the sensor of) the bomb: $|0\rangle$ switched-off,  $|1\rangle$ switched-on, as it actually happens in a CNOT gate. The incoming photon is represented by the target qubit, $q1$, in the state $|0\rangle$. If the sensor can ``detect the photon", i.e. the CNOT gate is working properly, the state of $q1$ changes to $|1\rangle$, producing 1 in the measuring unit (explosion). 
%The complete device can be thought as a means to determine whether a CNOT gate works, without actually turning it on.
%

Let us assume first that the bomb is a dud, i.e. the CNOT gate is fake and does not act in any way, independently of the state of the switch. Then, the first U2(0,0) gate prepares the switch state in an on-off superposition $(1/\sqrt{2})( |0\rangle + |1\rangle)$. (A Hadamard gate would do the same job).
%Then, the first Hadamard gate prepares the switch state in an on-off superposition. 
It plays a role similar to the beam splitter in the Mach-Zehnder interferometer. For its part,  the state of $q1$ does not change at any stage, as the CNOT gate does not work; so the measurement of $q1$ gives always 0 (no explosion). 
%This does not affect the state of $q1$ at any stage, as the CNOT gate does not work, so the measurement of $q1$ gives always '0' (no explosion). 
The action of the second U2(0,0) gate is to drive the control qubit into the $|1\rangle$ state. 
Its role is analogous to the second beam splitter in the Mach-Zehnder interferometer. At the end, the measurement of $q0$ will always give 1. 
%The action of the last two gates, $Z$ and $H$, is to drive the state of the switch into the $|1\rangle$ state. The role of these gates is analogous to the second beam splitter in the Mach-Zehnder interferometer. At the end, the measurement of $q0$ will always give '1'. 

Let us assume now that the bomb is not a dud. Then, the CNOT gate works properly and creates an entangled state, $(1/\sqrt{2}) (|00\rangle + |11\rangle)$, setting the bomb state in a superposition: exploding and non-exploding. If the measurement of $q1$ yields 1 (50\% of times), the "bomb explodes" and the CNOT gate has been actually switched on. If the result is 0 (50\% of times), the bomb has not been activated and the switch state, $q0$, also collapses at $|0\rangle$ (turned-off). Then the second U2(0,0) gate drives $q0$ to $(1/\sqrt{2}) (|0\rangle + |1\rangle)$. Thus the measurement of $q0$ will give 0 and 1 with equal probability. In total, 25\% of the times both the measurements of $q1$ and $q0$ yield 0. In those cases, the bomb has not been activated, but the result differs from that of a dud, which always gives 1 at $q0$. 
So a live bomb (CNOT gate) is selected without exploding (activating) it in the test.

Table \ref{tab:bombtester} shows the actual outcomes when the bomb tester circuits illustrated in Fig. \ref{fig:bombtester} are run in several IBM quantum computers \cite{IBMQ}. All of them show a good performance, except ibmq\_ourense and ibmq\_melbourne. The results for the corresponding dud circuits, also shown in the table, are close to the theoretical expectations.
In all cases we have re-designed the circuit to take profit of the qubits and connections with higher reliability. Every circuit has been run ten times at 8192 shots, in order to obtain the corresponding mean and the (unbiased) sample standard deviation, which are the values quoted in Table \ref{tab:bombtester}. These uncertainties capture not only the statistical fluctuations inherent to the quantum nature of the measurement, but also some of the systematic uncertainties associated to the specific performance of each quantum computer. The latter turn out to be sizeable and depend notably upon the timing of the execution.

We have used the same procedure throughout the paper.

\begin{widetext}

\vspace{0.2cm}
\begin{table}[h]
    \centering
    \begin{tabular}{c|c|c|c|c|c|}
       %circuit \textbackslash output
        & &
      00 & 10&  01& 11 \\
      \hline
      \hline 
      &theory &	{\bf 25} & $25$ & $25$ & $25$
      \\
      \cline{2-6}
     &ibmqx2 &	{\bf 24.8$\pm$0.6} & $27.9\pm 1.4$ & $22.50\pm0.5$ & $24.4\pm 0.4$
      \\
      Bomb&vigo &	{\bf27.0$\pm$0.9} & $26.4\pm 0.5$ & $26.8\pm 0.7$ & $20.1\pm 0.3$
      \\
      &ourense &	{\bf31.0$\pm$0.7} & $21.4\pm 0.5$ & $28.5\pm 0.5$ & $19.0\pm 0.7$
      \\
      &melbourne &	{\bf27.9$\pm$ 0.4} & $27.6\pm 0.6$ & $22.7\pm0.6$ & $21.9\pm 0.9$
      \\
      &johannesburg &	{\bf26.1$\pm$0.6} & $24.4\pm 0.4$ & $25.3\pm 0.4$ & $24.3\pm 0.4$
      \\
      \hline   
      \hline
         &theory &	$0$ & {\bf100} & $0$ & $0$
      \\
      \cline{2-6}
     &ibmqx2 &	$2.6\pm 0.2$ & {\bf97.0$\pm$0.2} & $0.05\pm 0.02$ & $0.4\pm 0.1$
      \\
      Dud&vigo &	$4.6\pm 0.2$ & {\bf94.9$\pm$0.2} & $0.05\pm 0.03$ & $0.5\pm 0.1$
      \\
      &ourense &	$2.7\pm0.2$ & {\bf95.2$\pm$0.3} & $0.06\pm 0.03$ & $2.0\pm 0.2$
      \\
       &melbourne &	$4.4\pm 0.3$ & {\bf95.0$\pm$0.3} & $0.06\pm 0.06$ & $0.6\pm 0.2$
      \\
      &johannesburg &	$2.7\pm 0.2$ & {\bf95.8$\pm$1.1} & $0.14\pm 0.14$ & $1.4\pm 1$
      \\
      \hline   
    \end{tabular}
    \caption{Theoretical and actual mean probability outcomes (in \%) of the bomb tester circuit of Fig. \ref{fig:bombtester} and the corresponding dud circuit run in several IBM quantum computers  after ten runs of 8192 shots. The names vigo, ourense, etc. are a shortening for ibmq\_vigo, ibmq\_ourense, etc.
    The first line indicates the possible outputs for the measurement of $q0$ and $q1$. The output 00 denotes the presence of an alive bomb, without exploding it; whereas 10 is the only possible output if the bomb is a dud.}
    \label{tab:bombtester}
\end{table}{}
\end{widetext}
%\vspace{0.2cm}
%\begin{table}[h]
%    \centering
%   \begin{tabular}{c|c|c|c|c|}
%       Result on $q0$ $q1$&
%      00 & 10&  01& 11 \\
%      \hline
%      \hline
%      theory &	$0.25$ & $0.25$ & $0.25$ & $0.25$
%      \\
%      ibmqx2 &	$0.269$ & $0.283$ & $0.221$ & $0.227$
%      \\
%      ibmq\_vigo &	$0.267$ & $0.237$ & $0.244$ & $0.252$
%      \\
%      \hline     
%    \end{tabular}
%    \caption{Theoretical and actual mean probability outcomes of the bomb tester circuit of Fig. \ref{fig:bombtester} run in two different IBM quantum computers.}
%    \label{tab:bombtester}
%\end{table}{}

\section{Counterfactual computations in a quantum computer}

Let us now build quantum circuits which implement simple counterfactual computations.

Note first that the circuit of Fig. \ref{fig:bombtester} can be indeed regarded as a circuit of that kind. Namely, the part of the circuit denoted as ``bomb" can be viewed as an (unknown to us) device that, when it is switched-on ($q0$ at $|1\rangle$), performs the computation on the input '0' (loaded in $q1$) yielding $|10\rangle\rightarrow |1r\rangle$ with $r=1$.
%, with $r$ unknown. 
Then, there is
%{\blu If we did not know what is the controlled-gate placed in the circuit and we wanted to know how does it operate on the input $|0\rangle$, there is}
a 25\% chance that in one run we determine that the output is indeed $r=1$, without actually switching on the device (measurements at $q0$ and $q1$ yielding 0). In addition, there is 50\% chance (when the measurement of $q1$ gives 1)  that the gate becomes turned on, and 25\% chance (when $q0$ and $q1$ yield 1 and 0 respectively) that  we cannot conclude anything . This matches the performance of the Jozsa counterfactual computation \cite{Jozsa:1998bq} (see steps 1-6 in section 1) for $N=2$, but with fewer operations and measurements.

Let us now construct quantum circuits which accomplish the Jozsa
procedure for arbitrary $N$. 
Recall that the method works for the case $r=1$,
which is the one we are going to implement. In that case, the switch and register states are perfectly represented by the control and target qubits of a CNOT gate.

\begin{figure}[htb]
\begin{center}
\includegraphics[height=4.0cm,clip=]{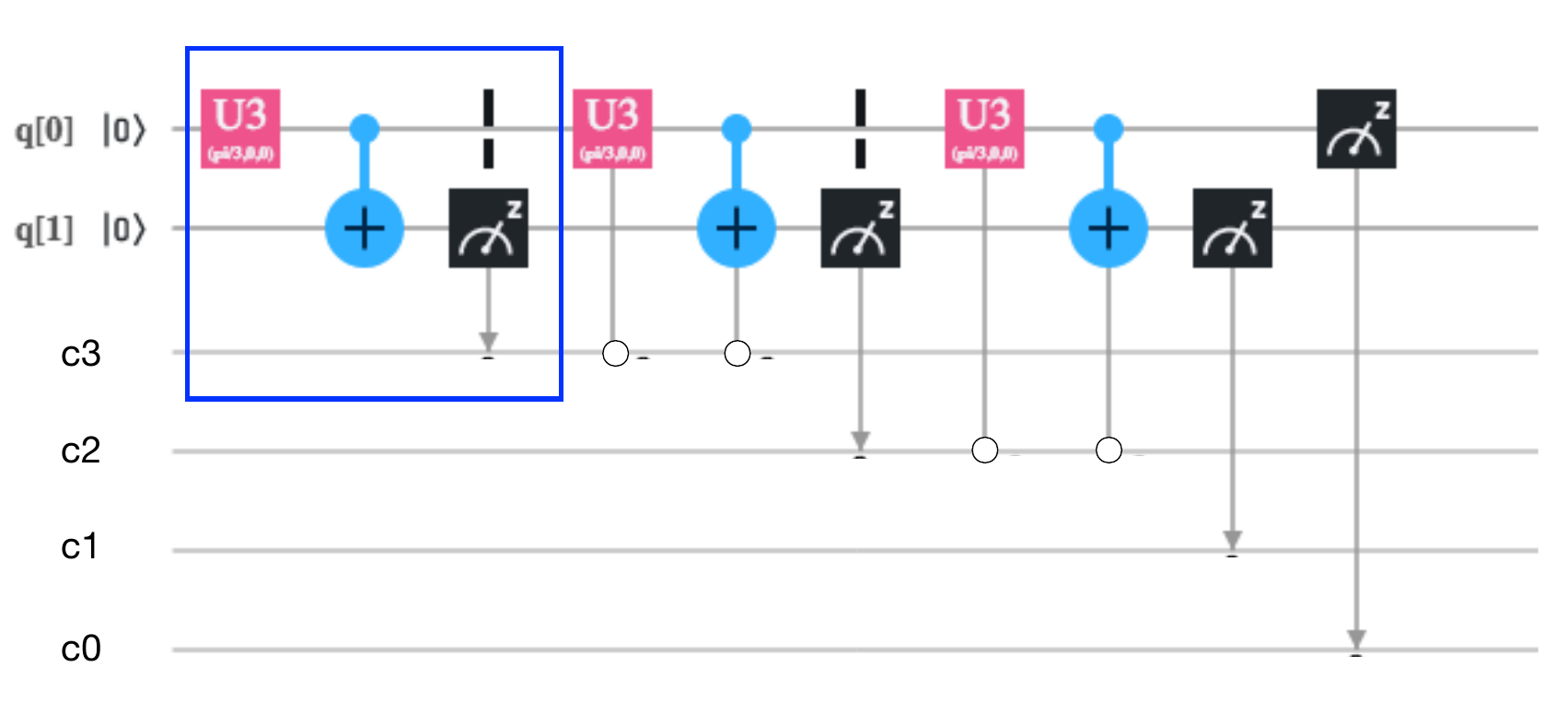} 
\caption{Circuit for the counterfactual computation proposed by Jozsa \cite{Jozsa:1998bq, Mitchison:1999uf} and described in points 1-6 of section I  (with $N=3$). The steps in the blue box are the ones to be repeated $N$ times.
}
\label{fig:sin_ancillas}
\end{center}
\end{figure}

Remember that the Jozsa procedure requires to perform intermediate measurements on $q1$, after which, if the measurement gives 0, the new state of the qubit is re-used as input; otherwise the procedure halts. This can be realized by the quantum circuit shown in Fig. \ref{fig:sin_ancillas}, consisting of 2 qubits and $N+1$ classical bits or `cbits' which save the results of the intermediate measurements (the circuit of the figure corresponds to $N=3$). The U3($\pi/N, 0,0$) gate in the circuit performs the $\theta-$rotation, see eq.(\ref{theta_rot}), while the CNOT gate is the (supposedly unknown) device that performs the calculation. Note that these gates are controlled by the classical bits and are only activated if the previous measurement on $q1$ yielded 0. This requirement is implemented by means of the IF operation, which is supported by the IBM quantum-computer simulator.
Hence, in theory, whenever one intermediate measurement on $q1$ gives 1, all the subsequent ones must yield 1 as well. These events are to be discarded. On the other hand, when {\em all} the cbits remain at 0, including the one associated with the $q0$ measurement, this signals that the result of the computation 
%of the (supposedly unknown) {\blu logic gate} acting on the input $|0\rangle$ produces the result $|r\rangle$ with 
is $r=1$. As discussed in section 1, this will occur with a probability $(\cos^2\theta)^{N}$, with $\theta=\frac{\pi}{2N}$. In contrast, if the result of the computation were $r=0$ (which corresponds to the same circuit replacing the CNOT operations by the identity), then all the cbits would remain at 0 with probability 1, except the one associated with the measurement of $q0$, which should become 1, an impossible output for $r=1$.

%The use of the IF operation (supported by the IBM quantum-computer simulator) allows to tag the cases when the result of the measurement on the second qubit yielded '1', which are to be discarded. This happens whenever any of the $c0-cN$ cbits becomes '1' at the end of the run. On the other hand, when all the cbits remain at '0', this signals that the computation of the (supposedly unknown) {\blu logic gate} acting on the input $|0\rangle$ produces the result $|r\rangle$ with $r=1$.

%

\begin{figure}[htb]
\begin{center}
\includegraphics[height=4.5cm,clip=]{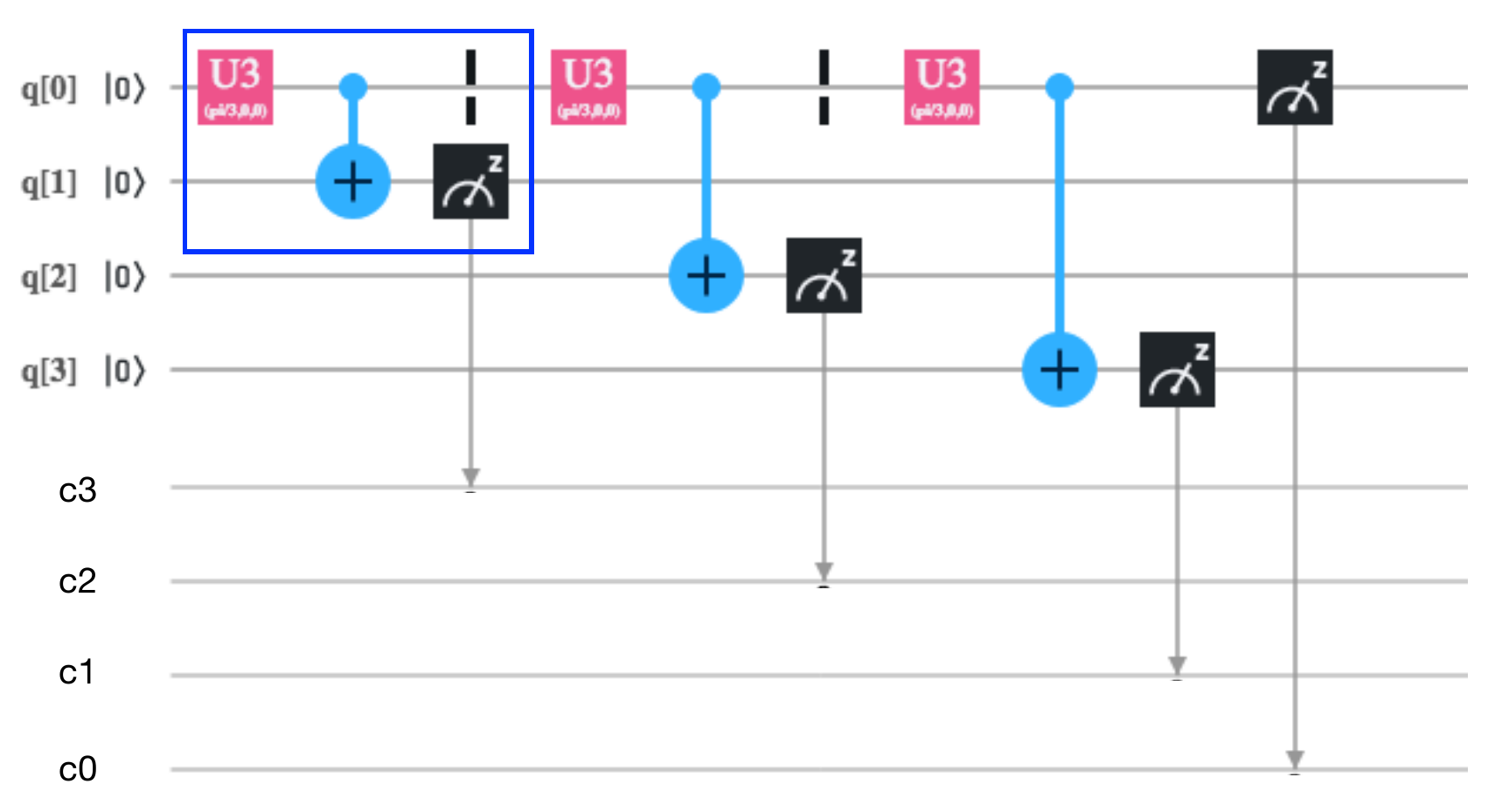}
\caption{Circuit for the counterfactual computation proposed by Jozsa, equivalent to that shown in Fig.\ref{fig:sin_ancillas}, but using ancillas. This circuit can be run in actual quantum computers. The figure corresponds to $N=3$, but it can be trivially extrapolated to any $N$ by repeating the steps in the blue box $N$ times.}
\label{fig:con_ancillas}
\end{center}
\end{figure}

Of course, when these circuits are run in the IBM quantum simulator, the results are in perfect agreement with the expectations. Unfortunately, the IF operation is not yet supported by the real IBM quantum computers. Still, we can create an equivalent circuit by using $N-1$ auxiliary qubits (ancillas). Fig. \ref{fig:con_ancillas} shows such circuit for $N=3$. The procedure is simply that after an intermediate measure on $q1$, this qubit is replaced by a new qubit prepared at $|0\rangle$ \footnote{This is equivalent to reset the $q1$ qubit at $|0\rangle$. However, this operation is not yet supported by the IBM quantum computer.} 
. Since the measurement destroys the possible entanglement between $q0$ and $q1$, this 
%does not modify in any way the state of the system. It 
is completely equivalent to re-using $q1$ when its measurement yielded 0 and thus it was reset at $|0\rangle$.
At the end of the procedure, the shots where all the cbits are at 0 are the successful ones. Again, this happens with a probability $(\cos^2\theta)^{N}$, with $\theta=\frac{\pi}{2N}$.

Tables \ref{tab:counterfactual} and \ref{tab:counterfactual2} show the theoretical and actual probabilities of success for the circuit of Fig. \ref{fig:con_ancillas} and 
several values of $N$, when run 
in various IBM quantum computers with different architectures: ibmqx2, ibmq\_vigo, ibmq\_ourense, ibmq\_melbourne and ibmq\_johannesburg \cite{IBMQ}. 
The latter two, with 15 and 20 qubits respectively, are the only ones which can cope with the $N>4$ circuits, since $N+1$ qubits are required in each case.
Again, in all instances we have designed the circuit to take profit of the qubits and connections with higher reliability. In addition, for each circuit, instead of using the automatic transpiling provided by IBM, we have re-designed
a ``pre-transpiled" circuit where all the connections among qubits actually exist in the corresponding computer's architecture. In this way we not only improve the performance, but, more importantly, we also eliminate the instability in the results caused by the randomness associated to the IBM's automatic transpiling procedure.  

%Once more, for $N\leq 4$ ibmq\_vigo shows a better performance than ibmqx2, while for  $N>4$, ibmq\_johannesburg gives systematically better results  than ibmq\_melbourne. \AC{Esto habrá que ver si es correcto finalmente. Además falta considerar Ourense}

Due to the accumulation of gates, one expects that, for increasing $N$, the departure from the theoretical predictions also increases; and this is typically the case. From Table \ref{tab:counterfactual},
For $N=2$ all q-computers, except ibmq\_ourense which shows a systematic excess, deliver a result compatible with the theoretical one within $\sim2$ standard deviations. Generally speaking, for $N\leq 4$
ibmq\_vigo shows the best performance, always within $\sim 3$ standard deviations.

%\BZ{Focusing first on the results for $N\leq 4$ from Table \ref{tab:counterfactual}, we observe that the ranking of performances across different q-computers varies depending on the value of $N$. For $N=2$ all q-computers but ibmq\_ourense deliver a result compatible with the theoretical one within $\sim2$ standard deviations. For $N=3$ ibmq\_vigo delivers the best results being compatible with the theory result at $\sim 3$ standard deviations, and for $N=4$ both ibmq\_vigo and ibmq\_ourense are compatible with the theory result at less than $\sim 1.6$ standard deviations.}
%
\vspace{0.2cm}
\begin{table}[h]
    \centering
    \begin{tabular}{r||c|c|c|}
       & 
      $N=2$ & $N=3$&  $N=4$ \\
      \hline
      \hline
      theory &	$25.0$ & $42.2$ & $53.1$ \\
      \hline
      %ibmqx2 &	$0.209$ & $0.326$ & $0.439$\\
      ibmqx2 &	$24.3\pm 0.3 $ & $38.0\pm 0.5$ & $47.8\pm 0.5$\\
      %ibmq\_vigo &	$0.275$ & $0.467$ & $0.550$\\
      vigo &	$24.8\pm 0.4$ & $46.3\pm 1.4$ & $52.0\pm 0.7$\\
      ourense & $32.4\pm 0.5$ & $48.1\pm 0.4$ & $53.3\pm 1.3$ \\
      melbourne & $24.9\pm 0.3$ & $35.2 \pm 3.3$ & $40.4\pm 4.2$\\
      johannesburg & $25.6\pm 0.6$ & $38.5\pm 1.2$ & $46.4\pm 0.6$
      \\
      \hline     
    \end{tabular}
    \caption{Theoretical and actual mean probability outcomes (in \%) of the all-zero 
    %event
    output for the circuit of Fig. \ref{fig:con_ancillas} and  $N\leq 4$, run in several IBM quantum computers.
    }
    \label{tab:counterfactual}
\end{table}{}
For $N>4$ we see from Table \ref{tab:counterfactual2}, ``uncorr." columns, that the performance gets worse, as expected; although ibmq\_johannesburg shows better results in general than ibmq\_melbourne, in part due to its richer connectivity.
Hence, for these cases we have implemented a simple error mitigation procedure, dealing exclusively with the errors in the measurements. Namely, for each circuit we extract the readout error simply by running in a row the same circuit with all gates removed (hereafter referred to as the ``calibration circuit") and counting the final percentage of  $0...0$ outputs, which in theory should be 100\%. 
Then we apply the inverse of this factor to the original result, obtaining the final corrected value quoted in the ``corr." columns of Table \ref{tab:counterfactual2}. Note that this procedure is appropriate in this case since for $N>4$ the theoretical probability to obtain an output with all 0s except one 1, e.g. $10...0$, is very small, namely $(\sin\frac{\pi}{2N})^4(\cos\frac{\pi}{2N})^{2(N-2)}$.
Hence, the total number of erroneous counts in which that '1' is flipped and thus we read $00...0$ is negligible (the flip of two or more 1s is even more unlikely).
%Consequently the probability that the system counts erroneously one of those outputs as 0...0 is extremely small (it requires the flip of two individual outcomes). 
Thus all the relevant leaking of probability due to errors in the measurement goes essentially from the $0...0$ output to the others and not the other way around; and it is well estimated by the calibration circuit \footnote{Alternatively, one can use the readout errors for the different qubits provided by the IBM Quantum Experience platform everyday \cite{IBMQ}. The result is similar albeit less accurate.
}. 
The uncertainties quoted in the ``corr." columns of Table \ref{tab:counterfactual2} correspond to the combination of those associated to the counterfactual and calibration circuits, according to the standard uncertainty propagation techniques. After this error correction the results improve appreciably, at least for ibmq\_johannesburg, even though for $N>7$ they are distant from the theoretical expectations.

%We have implemented the $N>4$ cases in the 15-qubit ibmq\_16\_melbourne as well as in the 20-qubit ibmq\_johannesburg quantum computers, but the results are very poor are unstable, far from the theoretical expectations, even after optimizing the configuration of the circuit to take profit of the qubits and connections with higher reliability. Hence the results are not worth to be presented. Our conclusion here is that the ibmq\_16\_melbourne quantum computer is working at the moment with much less reliability than the other IBM quantum computers.
\begin{widetext}
\begin{center}
\vspace{0.2cm}
\begin{table}[h]
    \centering
    \begin{tabular}{c||c||c|c||c|c|}
 & theory & \multicolumn{2}{c}{melbourne $~~$} & 
    \multicolumn{2}{c}{johannesburg} \\
    \hline
    & $~$ & uncorr. & \textbf{corr.} & uncorr. & \textbf{ corr.} \\
 \hline\hline
 $N=5$ & 60.5 & $48.1\pm 0.8$ & $50.2\pm 1.0$ & $49.8\pm 0.5$ & $56.1\pm 1.0$ \\
 $N=6$ & 66.0 & $49.6\pm 1.5$ & $51.9\pm 1.6$ & $47.2\pm 1.4$ & $56.7\pm 3.1$ \\
 $N=7$ & 70.1 & $40.6\pm 2.5$ & $44.6\pm 3.5$ & $45.3\pm 3.0$ & $54.5\pm 4.1$ \\
 $N=8$ & 73.3 & $30.4\pm 0.8$ & $34.7\pm 1.0$ & $45.3\pm 1.3$ & $55.3\pm 1.9$ \\
 $N=9$ & 75.9 & $26.6\pm 0.5$ & $31.3\pm 0.7$ & $37.2\pm 3.8$ & $55.6\pm 6.7$ \\
 \hline
    \end{tabular}
    \caption{Theoretical and actual mean probability outcomes (in \%) of the all-zero 
    %event 
    output for the circuit of Fig. \ref{fig:con_ancillas} and several values of $N>4$, run in two different IBM quantum computers. The ``uncorr" (``corr") columns correspond to the results before (after) implementing the simple error-correction procedure described in the text. 
    }
    \label{tab:counterfactual2}
\end{table}{}
\end{center}
\end{widetext}

\section{Conclusions}

 A fascinating feature of quantum mechanics is the possibility of realizing interaction-free (also called counterfactual) measurements, in which non-trivial information about a system is obtained without disturbing it. The concept has been also applied to show the the theoretical possibility of counterfactual computations, in which a (typically simple) computation is realized with the computer switched off. 
 
In this paper we have shown how to implement both effects in a quantum computer by using simple quantum circuits. More specifically, following the spirit of the Elitzur-Vaidam experiment \cite{Elitzur:1993xh}, the simple quantum circuit of Fig. \ref{fig:bombtester} allows to select a ``live bomb" (represented by a live CNOT gate) without exploding (activating) it with a 25\% probability. We have run the circuit in several IBM quantum computers, obtaining results close to the theoretical expectations.

Concerning counterfactual computations, we have designed quantum circuits that implement the Jozsa protocol \cite{Jozsa:1998bq, Mitchison:1999uf} for a simple counterfactual computation. This protocol gives the possibility to obtain the result of a simple  1-bit to 1-bit computation, namely $f(0)=1$, without actually switching on the computer that performs it, with a $(\cos^2\theta)^{N}$ probability, where $N$ is the number of iterations of the protocol. 
%More specifically, the circuit of Fig. \ref{fig:con_ancillas} implements such protocol and can be implemented in IBM quantum computers.
For each value of $N$ we have built the corresponding circuit (illustrated in Fig. 4 for $N=3$) that implements such protocol and can be ran in the IBM q-computers.

For $N\leq4$ the results are close to theoretical expectations in most of the q-computers probed. As $N$ increases, the departure from the theoretical predictions also increases due
to the accumulation of gates. For $N>5$ we have implemented a simple procedure which mitigates the error due to the measurement and provides a perceptible improvement of the results.

\section{Acknowledgements}

%This work is based on a master thesis presented by D.G.M. at the master in Theoretical Physics of the Universidad Autónoma de Madrid (UAM) during the course 2016-2017. 
We thank E. L\'opez and G. Sierra for inspiring conversations and advise. We also thank
the IBM Quantum team for making multiple devices available
 via the IBM Quantum Experience. The access to the IBM Quantum Experience has been provided by the
CSIC IBM Q Hub.  
We acknowledge the SEV-2016-0597 of the Centro de Excelencia Severo Ochoa Programme. B.Z. is further supported by the Programa
Atracci\'on de Talento de la Comunidad de Madrid under grant n. 2017-T2/TIC-5455,
from the Comunidad de Madrid/UAM ``Proyecto de J\'ovenes Investigadores'' grant n. SI1/PJI/2019-00294, from Spanish ``Proyectos de I+D de Generaci\'on de Conocimiento'' via
grants PGC2018-096646-A-I00 and PGC2018-095161-B-I00.

%The views expressed  are those of the authors and do not reflect the official policy or position of IBM or the IBM Quantum Experience.

  \end{document}